\documentclass[12pt,preprint]{aastex}

\newcommand{\himpc}{{\hbox {$h^{-1}$}{\rm Mpc}}}
\newcommand{\iint}{{\int\!\!\!\!\int}}
\newcommand{\simgt}{\lower.5ex\hbox{$\; \buildrel > \over \sim \;$}}
\newcommand{\simlt}{\lower.5ex\hbox{$\; \buildrel < \over \sim \;$}}
\slugcomment{RESCEU-2/01\quad UTAP-387/2001}
\shorttitle{Log-normal PDF for Cosmological Density Fluctuations}
\shortauthors{Kayo, Taruya, \& Suto}
\begin{document}
%
\title{Probability Distribution Function of Cosmological Density
 Fluctuations from Gaussian Initial Condition:\\ Comparison of
One- and Two-point Log-normal Model Predictions with
N-body Simulations
}
%
\author{Issha Kayo, Atsushi Taruya, and Yasushi Suto}
\affil{Department of Physics and Research Center for
    the Early Universe (RESCEU), \\ School of Science, University of
    Tokyo, Tokyo 113-0033, Japan.}
\email{kayo@utap.phys.s.u-tokyo.ac.jp, 
ataruya@utap.phys.s.u-tokyo.ac.jp, suto@phys.s.u-tokyo.ac.jp}
%
\received{2001 April 2}
\accepted{2001 ??}
\begin{abstract}
 We quantitatively study the probability distribution function (PDF) of
 cosmological nonlinear density fluctuations from N-body simulations
 with Gaussian initial condition.  In particular, we examine the
 validity and limitations of one-point and two-point log-normal PDF
 models against those directly estimated from the simulations.  We find
 that the one-point log-normal PDF describes very accurately the
 cosmological density distribution even in the nonlinear regime (the rms
 variance $\sigma_{\rm nl} \simlt 4$ and the over-density $\delta \simlt
 100$).  Furthermore the two-point log-normal PDFs are also in good
 agreement with the simulation data from linear to fairly nonlinear
 regime, while slightly deviate from them for $\delta \simlt -0.5$.
 Thus the log-normal PDF can be used as a useful empirical model for the
 cosmological density fluctuations. While this conclusion is fairly
 insensitive to the shape of the underlying power spectrum of density
 fluctuations $P(k)$, models with substantial power on large scales,
 i.e., $n\equiv d\ln P(k)/d \ln k \simlt -1$, are
 better described by the log-normal PDF.  On the other hand, we note
 that the one-to-one mapping of the initial and the evolved density
 fields consistent with the log-normal model does not approximate the
 broad distribution of their mutual correlation even on average. Thus
 the origin of the phenomenological log-normal PDF approximation still
 remains to be understood.
\end{abstract} 
%
%
\section{INTRODUCTION}
\label{sec:intro}

Probability distribution function (PDF) of the cosmological density
fluctuations is the most fundamental statistic characterizing the
large-scale structure of the universe.  In the standard picture of
gravitational instability, the PDF of the primordial density
fluctuations which are responsible for the current structures in the
universe is assumed to obey the random-Gaussian. Therefore it is fully
specified by the two-point correlation function $\xi(r)$, or
equivalently, the power spectrum $P(k)$.  As long as the density
fluctuations are in the linear regime, their PDF remains Gaussian.  Once
they reach the nonlinear stage, however, their PDF significantly
deviates from the initial Gaussian shape due to the strong non-linear
mode-coupling and the non-locality of the gravitational dynamics.  The
functional form for the resulting PDFs in nonlinear regimes are not
known exactly, and a variety of phenomenological models have been
proposed (Saslaw 1985; Suto, Itoh, \& Inagaki 1990; Lahav et
al. 1993; Gazta\~{n}aga \& Yokoyama 1993; Suto 1993; Ueda \& Yokoyama
1996). 
Once such
one-point PDF is specified, one can characterize the clustering of the
universe with the higher-order statistics like skewness and kurtosis.
Moreover the two-point PDF is useful in estimating the errors in the
one-point statistics due to the finite sampling since the measurement at
different positions is not independent and their correlations are
supposed to be dominated by the two-point correlation function (Colombi,
Bouchet, \& Schaeffer 1995; Szapudi \& Colombi 1996).  Also the
two-point PDF plays an important role in analytical modeling of dark
halo biasing on two-point statistics.

From an empirical point of view, Hubble (1934) first noted that the
galaxy distribution in angular cells on the celestial sphere may be
approximated by a log-normal distribution, rather than a Gaussian.  More
recent analysis of the three-dimensional distribution of galaxies indeed
confirmed this (e.g., Hamilton 1985; Bouchet et al. 1993; Kofman et
al. 1994). Interestingly, several N-body simulations in cold dark matter
(CDM) models also indicated that the PDF of density fluctuations is
fairly well approximated by the log-normal (e.g., Coles \& Jones 1991;
Kofman et al. 1994; Taylor \& Watts 2000), at least in a weakly
nonlinear regime.

Those observational and numerical indications have not yet been
understood theoretically; Bernardeau (1992, 1994) showed that the PDF
computed from the perturbation theory in a weakly nonlinear regime
approaches the log-normal form only when the primordial power spectrum
is proportional to $k^{n}$ with $n=-1$. On the basis of this result,
Bernardeau \& Kofman (1995) argued that the successful fit of the
log-normal PDF in the CDM models should be interpreted as accidental,
and simply resulted from the fact that those model have the power
spectrum well approximated by $k^{n_{\rm eff}}$ with $n_{\rm eff} \simeq
-1$ on scales of cosmological interest. They claimed that the log-normal
PDF may fail either in a highly nonlinear regime or in models with power
spectrum with $n_{\rm eff}\neq -1$.  In fact, Ueda \& Yokoyama (1996)
conclude that the log-normal PDF does not fit well the PDF in a highly
nonlinear regime, from the analysis of CDM simulations by Suginohara \&
Suto (1991) employing $N=64^3$ particles in a $100$Mpc box.

The aim of this paper is to study the extent to which the log-normal
model describes the PDF in weakly and highly nonlinear regimes using the
high-resolution N-body simulations with $N=256^3$.  In particular, we
extend our analysis to the two-point PDF, in addition to the one-point
PDF discussed previously.  
Bernardeau (1996) analytically computed the two-point PDF using the
perturbation technique and compared somewhat indirectly with N-body
simulations in a weakly nonlinear regime.  
In contrast, we focus on the highly nonlinear
regime, and examine the validity of the empirical log-normal model.

This paper is organized as follows.  Section \ref{sec:lognormal}
describes the log-normal PDF derived through the one-to-one mapping between
the linear and nonlinear density fields.  The detailed comparison
between the log-normal predictions and N-body results is presented in \S
\ref{sec:comparison}.  Finally \S \ref{sec:conclusion} is devoted to
conclusions and discussion.

\section{PROBABILITY DISTRIBUTION FUNCTIONS 
FROM THE LOG-NORMAL TRANSFORMATION}
\label{sec:lognormal}

In this section we briefly outline the derivation of the log-normal
PDF assuming the one-to-one corresponding between the linear and
evolved density fluctuations.  Throughout the paper, we consider the
mass density field, $\rho(\mbox{\boldmath$x$};R)$ at the position
$\mbox{\boldmath$x$}$ smoothed over the scale $R$. This is related to
the {\it unsmoothed} density field $\rho(\mbox{\boldmath$x$})$ as
\begin{eqnarray}
  \label{eq: smoothing}
  \rho(\mbox{\boldmath$x$}; R) &=& \int d^{3} \mbox{\boldmath$y$} 
  W(|\mbox{\boldmath$x$}-\mbox{\boldmath$y$}|;R)\,\,
\rho(\mbox{\boldmath$y$}) 
\nonumber \\
 &=& \int \frac{d^{3}\mbox{\boldmath$k$}}{(2\pi)^3} \tilde{W}(kR) 
\tilde{\rho}(\mbox{\boldmath$k$}) \,e^{-i\mbox{\boldmath$k$}\cdot
\mbox{\boldmath$x$}}.
\end{eqnarray}
In the above expression, $W$ denotes the window function, and
$\tilde W$ and $\tilde\rho$ represent the Fourier transforms of the
corresponding quantities. In what follows, we adopt the two
conventional windows:
\begin{equation}
 \tilde{W}(x)=
 \cases{
  e^{-x^2/2} & (Gaussian),\cr
  3(\sin x-x\cos x)/x^3 & (Top-hat).\cr
  }  \label{eq: window_func}
\end{equation}
Then the density contrast at the position $\mbox{\boldmath$x$}$ is
defined as $\delta(\mbox{\boldmath$x$};R)
\equiv(\rho(\mbox{\boldmath$x$};R)-\bar\rho)/\bar\rho$, with
$\bar\rho$ denote the spatial average of the smoothed mass density
field. For simplicity we use $\delta$ to denote
$\delta(\mbox{\boldmath$x$};R)$ unless otherwise stated.

\subsection{One-point Log-normal PDF}
\label{subsec:1pLN}
The one-point log-normal PDF of a field $\delta$ is defined as
\begin{eqnarray}
 P^{(1)}_{\rm LN}(\delta) =
  \frac{1}{\sqrt{2\pi\sigma_1^2}}\exp\left[ -\frac{\{\ln(1+\delta) +
 \sigma_1^2/2\}^2}{2\sigma_1^2}\right]\frac{1}{1+\delta} .
\label{eq:1pLNPDF}
\end{eqnarray}
The above function is characterized by a single parameter $\sigma_1$
which is related to the variance of $\delta$. Since we use $\delta$ to
represent the density fluctuation field smoothed over $R$, its
variance is computed from its power spectrum $P_{\rm nl}$ explicitly as
\begin{equation}
 \sigma_{\rm nl}^2(R)
  \equiv
  \frac{1}{2\pi^2}\int_0^\infty P_{\rm nl}(k)\tilde{W}^2(kR)k^2 dk .
\end{equation}
Here and in what follows, we use subscripts ``lin'' and ``nl'' to
distinguish the variables corresponding to the primordial (linear)
and the evolved (nonlinear) density fields, respectively. Then
$\sigma_1$ depends on the smoothing scale $R$ alone and is given by
\begin{eqnarray}
 \sigma_1^2(R) = \ln\left[1+\sigma_{\rm nl}^2(R)\right]. 
\end{eqnarray}
Given a set of cosmological parameters, one can compute $\sigma_{\rm
nl}(R)$ and thus $\sigma_1(R)$ very accurately using a fitting formula
for $P_{\rm nl}(k)$ (e.g., Peacock \& Dodds 1996, hereafter PD).  In
this sense, the above log-normal PDF is completely specified and
allows the definite comparison against the numerical simulations (\S
3).

It is known that the above log-normal function may be obtained from
the one-to-one mapping between the linear random-Gaussian and the
nonlinear density fields (e.g., Coles \& Jones 1991).  Define a linear
density field $g$ smoothed over $R$ obeying the Gaussian PDF:
\begin{eqnarray}
 P^{(1)}_{\rm G}(g) =
\frac{1}{\sqrt{2\pi\sigma_{\rm lin}^2}}
\exp\left(-\frac{g^2}{2\sigma_{\rm lin}^2}\right),
\end{eqnarray}
where the variance is computed from its linear power spectrum:
\begin{eqnarray}
 \sigma_{\rm lin}^2(R) \equiv
\frac{1}{2\pi^2}\int_0^\infty P_{\rm lin}(k)\tilde{W}^2(kR)k^2dk .
\end{eqnarray}
If one introduces a new field $\delta$  from $g$ as
\begin{equation}
 1+\delta = \frac{1}{\sqrt{1+\sigma_{\rm nl}^2}}
  \exp\left\{\frac{g}{\sigma_{\rm lin}}
\sqrt{\ln(1+\sigma_{\rm nl}^2)}\right\},
  \label{eq:trans1}
\end{equation}
the PDF for $\delta$ is simply given by $(dg/d\delta)P^{(1)}_{\rm
G}(g)$ which reduces to equation (\ref{eq:1pLNPDF}). 

At this point, the transformation (\ref{eq:trans1}) is nothing but a
mathematical procedure to relate the Gaussian and the log-normal
functions. Thus there is no physical reason to believe that the new
field $\delta$ should be regarded as a nonlinear density field evolved
from $g$ even in an approximate sense. In fact it is physically
unacceptable since the relation, if taken at face value, implies that
the nonlinear density field is completely determined by its linear
counterpart locally. We know, on the other hand, that the nonlinear
gravitational evolution of cosmological density fluctuations proceeds
in a quite nonlocal manner, and is sensitive to the surrounding mass
distribution.

Nevertheless the fact that the log-normal PDF provides a good fit to
the simulation data empirically as discussed in \S 1 implies that the
transformation (\ref{eq:trans1}) somehow captures an important aspect
of the nonlinear evolution in the real universe. In \S 3, we present
detailed discussion on this problem. Before that, we derive the
two-point log-normal PDF by applying this transformation in the next
subsection.

\subsection{Two-point Log-normal PDF}
\label{subsec:2pLN}

Consider two density fields, $\delta_1=\delta(\mbox{\boldmath$x$}_1;
R)$ and $\delta_2=\delta(\mbox{\boldmath$x$}_2; R)$, located at
$\mbox{\boldmath$x$}_1$ and $\mbox{\boldmath$x$}_2$ smoothed over $R$.
We denote the two-point PDF by $P^{(2)}(\delta_1, \delta_2; r)$ for
the two fields with a specified separation $r$, i.e., satisfying the
condition $|\mbox{\boldmath$x$}_1-\mbox{\boldmath$x$}_2|=r$.

In the case of the Gaussian fields $g_1$ and $g_2$, this two-point PDF
is given by the bi-variate Gaussian (e.g., Bardeen et al. 1986):
\begin{equation}
 P^{(2)}_{\rm G}(g_1, g_2; r)
  =\frac{1}{2\pi\sqrt{\det M}}\exp\left\{-\frac12(g_1, g_2)M^{-1}
   \left( \begin{array}{cc}g_1\\g_2\end{array}\right)\right\},
\end{equation}
where
\begin{equation}
 M \equiv
  \left(\begin{array}{cc} \langle g_1^2\rangle & \langle g_1 g_2\rangle 
\\ 
\langle g_1 g_2\rangle  & \langle g_2^2\rangle \end{array}\right)\\
 = \left(\begin{array}{cc} \sigma_{\rm lin}^2 & \xi_{\rm lin}(r)\\
   \xi_{\rm lin}(r) & \sigma_{\rm lin}^2\end{array}\right),
\end{equation}
and
\begin{equation}
 \xi_{\rm lin}(r; R) =  \frac{1}{2\pi^2}\int_0^\infty 
P_{\rm lin}(k)\tilde{W}^2(kR)\frac{\sin(kr)}{kr}k^2 dk.
\end{equation}

From an analogy of equation (\ref{eq:trans1}), let us assume that the
transformation from $(g_1, g_2)$ to $(\delta_1, \delta_2)$ is given by
the form:
\begin{equation}
 1+\delta_i = \alpha e^{\beta g_i} . \quad (i=1, 2)\label{eq:trans2}
\end{equation}
The coefficients $\alpha$ and $\beta$ are determined by the following
conditions:
\begin{equation}
 \langle\delta_1\rangle = \langle\delta_2\rangle = 0,
\end{equation}
\begin{equation}
 \langle\delta_1^2\rangle = \langle\delta_2^2\rangle =\sigma_{\rm nl}^2,
\end{equation}
\begin{equation}
 \langle\delta_1\delta_2\rangle = \xi_{\rm nl}(r) .
\end{equation}
In the above expressions, the angular bracket denotes the average over
the two-point PDF which in the present model, reduces to
\begin{eqnarray}
 \langle{\cal F}(\delta_1, \delta_2) \rangle &\equiv&
  \iint_{-1}^\infty {\cal F}(\delta_1, \delta_2) P^{(2)}(\delta_1,
  \delta_2; r) d\delta_1 d\delta_2 \cr 
&=& \iint_{-\infty}^\infty  {\cal F}
(\delta_1(g_1), \delta_2(g_2)) P_{\rm G}(g_1, g_2; r) dg_1 dg_2 .
\end{eqnarray}
After a straightforward calculation, one obtains
\begin{equation}
 \alpha = \frac{1}{\sqrt{1+\sigma^2_{\rm nl}}}, \quad
 \beta = \sqrt{\frac{\ln(1+\xi_{\rm nl})}{\xi_{\rm lin}}} .
\end{equation}
Then this procedure yields the two-point log-normal PDF:
\begin{equation}
 P^{(2)}_{\rm LN}(\delta_1, \delta_2; r) =
 \frac{1}{2\pi\sqrt{S^2-X^2}}
 \exp\left[ - \frac{S\{ L_1^2 + L_2^2\} - 2X L_1L_2}{2\{ S^2
 - X^2\}}\right]\frac{1}{(1+\delta_1)(1+\delta_2)}, \label{eq:2pLN}
\end{equation}
where
\begin{eqnarray}
 X&\equiv& \ln(1+\xi_{\rm nl}),\\
 S&\equiv& \ln (1+\sigma_{\rm nl}^2),\\
 L_i&\equiv& \ln\left\{(1+\delta_i)\sqrt{1+\sigma_{\rm nl}^2}\right\},
\quad (i=1, 2).
\end{eqnarray}
Again the nonlinear two-point correlation function $\xi_{\rm nl}(r;
R)$ can be computed as
\begin{equation}
 \xi_{\rm nl}(r; R) =
  \frac{1}{2\pi^2}\int_0^\infty 
P_{\rm nl}(k)\tilde{W}^2(kR)\frac{\sin(kr)}{kr}k^2 dk ,
  \label{eq:xi_PD}
\end{equation}
and thus equation (\ref{eq:2pLN}) can be fully specified using the PD
nonlinear power spectrum.

\section{THE LOG-NORMAL PDFS AGAINST N-BODY SIMULATIONS}
\label{sec:comparison}

The previous section discussed a prescription to derive one-point and
two-point log-normal PDFs assuming the one-to-one mapping between the
linear and the nonlinear density fields. As remarked, however, the
assumption does not seem to be justified in reality.  So in this
section we compare the log-normal PDFs extensively with the results of
cosmological N-body simulations, and discuss their validity and
limitations. The analysis for the one-point PDF below significantly
increases the range of $\delta$ compared with several previous work.
As far as we know, the direct estimation of the two-point PDF in the
nonlinear regime from simulations has not been performed before and this
is the first attempt.

\subsection{N-body Simulations}
\label{subsec:sim}

For the present analysis, we use a series of cosmological N-body
simulations in three CDM models (SCDM, LCDM and OCDM for Standard,
Lambda and Open CDM models, respectively; Jing \& Suto 1998) and four
scale-free models with the initial power spectrum $P(k)\propto k^n$
(n=1, 0, $-1$, and $-2$; Jing 1998).  All the models employ $N=256^3$
dark matter particles in a periodic comoving cube $L_{\rm box}^3$, and
are evolved using the P$^3$M code.  The gravitational softening length
is $L_{\rm box}/2560$ ($3L_{\rm box}/5120$) for the CDM (scale-free)
models, and kept fixed in the comoving length.  The amplitude of the
fluctuations in CDM model , $\sigma_8$, is normalized according to the
cluster abundance (e.g., Kitayama \& Suto, 1997).  The scale-free models
assume the Einstein-de Sitter universe (the density parameter
$\Omega_0=1$, the cosmological constant $\lambda_0=0$).  The other
parameters of the CDM models are summarized in Table \ref{tbl:simpara}.

The mass density fields are computed on $512^3$ grids in the
simulation box. First we assign particles to each grid point using the
cloud-in-cell interpolation. Then we apply the smoothing kernel in the
Fourier space, and then obtain the {\it smoothed} density fields after
the inverse Fourier transform. Note that the density fields on those
grids are not completely independent for $R>L_{\rm box}/512$, and we
do heavy over-sampling in this sense. Nevertheless the error-bars
quoted in our results below are estimated from the variance among the
three different realizations for each model (except the $n=-1$ model
which has two realizations only), and thus are free from the
over-sampling.

\subsection{The one-point PDF}
\label{subsec:1pcompare}

\begin{figure}
\plotone{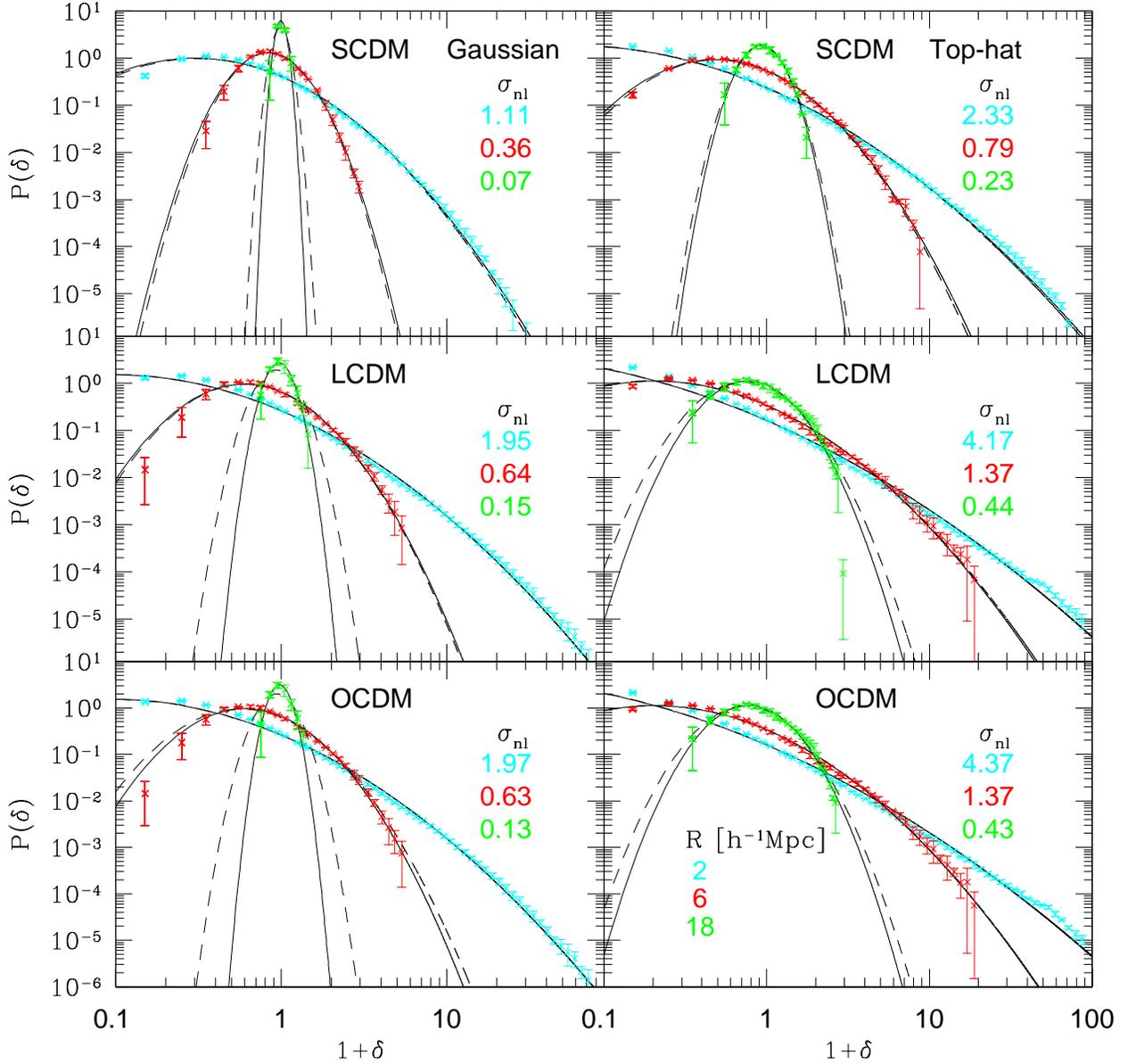} 
\figcaption{One-point PDFs in CDM models with Gaussian ({\it left}
 panels) and top-hat ({\it right} panels) smoothing windows; $R=2$\himpc
 ({\it cyan}), 6\himpc ({\it red}), and 18\himpc ({\it green}). The top,
 middle and bottom panels correspond to the PDFs in SCDM, LCDM, and OCDM.
 The {\it solid} and {\it long-dashed} lines represent the log-normal PDF
 adopting $\sigma_{\rm nl}$ calculated directly from the simulations and
 estimated from the nonlinear fitting formula of Peacock \& Dodds (1996),
 respectively.
 The values of $\sigma_{\rm nl}$ in each panel are estimated from the
 simulations.
 \label{fig:1pLN_CDM}}
\end{figure}

Consider first the one-point PDFs in CDM models
(Fig. \ref{fig:1pLN_CDM}).  The PDFs are constructed by binning the
data with $\Delta \delta = 0.1$, but we do not plot all the data
points just for an illustrative purpose. We compute the density
fields smoothed over Gaussian ({\it Left panels}) and Top-hat ({\it
Right panels}) windows with different smoothing
lengths;$R=2h^{-1}$Mpc, $6h^{-1}$Mpc and $18h^{-1}$Mpc plotted in
cyan, red and green symbols with error-bars, respectively.  The
corresponding values of $\sigma_{\rm nl}$ are summarized in Table
\ref{tbl:sigma_cdm}, and also shown on each panel.  Solid lines show
the log-normal PDFs adopting those $\sigma_{\rm nl}$ directly
evaluated from simulations. The agreement between the log-normal model
and the simulation results is quite impressive. A small deviation is
noticeable only for $\delta \simlt -0.5$.

We also show the log-normal PDFs in dashed lines, adopting $\sigma_{\rm
nl}$ calculated from the nonlinear fitting formula of PD (values in
parenthesis in Table \ref{tbl:sigma_cdm}).  Therefore the predictions do
not use the specific information of the current simulations, and are
completely independent in this sense.  While these predictions are in
good agreement with simulation data for $R \simlt 6h^{-1}$Mpc, the
results for $R=18h^{-1}$Mpc are rather different. Actually this
discrepancy should be ascribed to the simulations themselves, not to the
model predictions; Table \ref{tbl:sigma_cdm} indicates that the
$\sigma_{\rm nl}$ in the current simulations become systematically
smaller than the PD predictions for larger $R$.  This is because the
simulations assume (incorrectly) no fluctuations beyond the scale of the
simulation boxsize $L_{\rm box}$. This constraints systematically reduce
the fluctuations as the smoothing scale $R$ approaches $L_{\rm box}$.
We made sure that this is indeed the case by repeating the same analysis
using the CDM simulations evolved in $L_{\rm box}=300 h^{-1}{\rm Mpc}$
(Jing \& Suto 1998); the variance of fluctuations at $R \leq 18
h^{-1}$Mpc from the simulations agrees with the PD prediction within 2\%
accuracy.  Thus we conclude that the log-normal PDF with the PD formula
reproduces accurately the simulation results in the CDM models.

\begin{figure}
\plotone{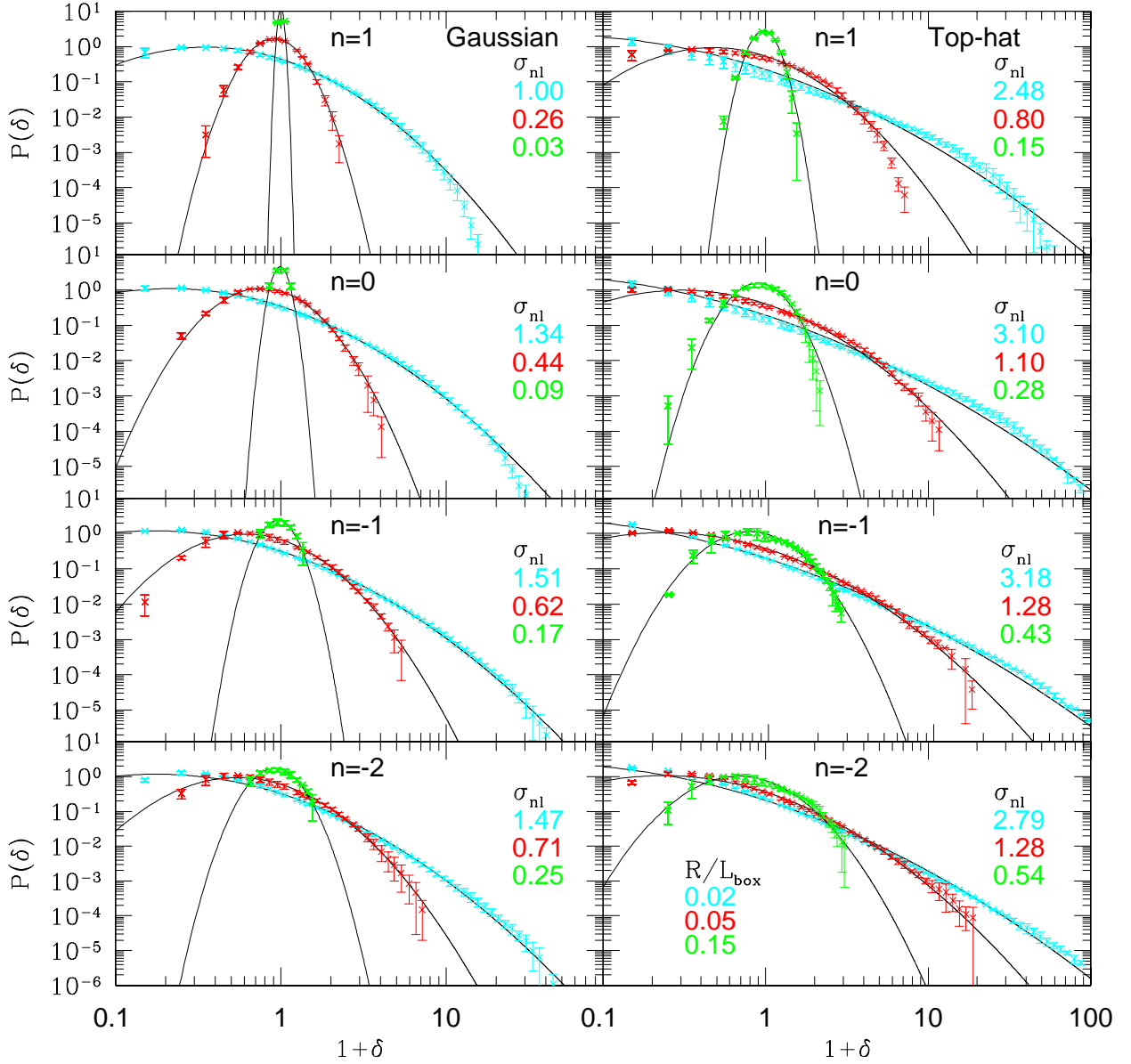}
\figcaption{Same as Fig \ref{fig:1pLN_CDM}, but in the scale-free models
 ($n=1$ to $-2$ from {\it top} to {\it bottom}); $R=0.02L_{\rm box}$
 ({\it cyan}), $0.05L_{\rm box}$ ({\it red}), and $0.15L_{\rm box}$
 ({\it green}).  \label{fig:1pLN_SF}}
\end{figure}

Next turn to the scale-free models.  Figure \ref{fig:1pLN_SF} shows
the similar plots corresponding to Figure \ref{fig:1pLN_CDM} but for
$n=1$ to $-2$ models (from top to bottom panels).
In this figure, we compare the simulation data (symbols with error-bars)
with the log-normal PDF predictions (solid lines) adopting $\sigma_{\rm
nl}$ from simulations (Table \ref{tbl:sigma_sf}).
Generally their agreement is good also in these models.
A closer look at Figure \ref{fig:1pLN_SF}, however, reveals that the
simulation results start to deviate from the log-normal predictions at
both high and low density regions, and that the deviation seems to
systematically increase as $n$ becomes larger.
While this tendency is qualitatively consistent with the earlier claim
by Bernardeau (1994) and Bernardeau \& Kofman (1995) on the basis
of the perturbation theory, our fully nonlinear simulations show that
the deviation from the log-normal PDF is not so large even in these
scale-free models.

\begin{figure}
\plotone{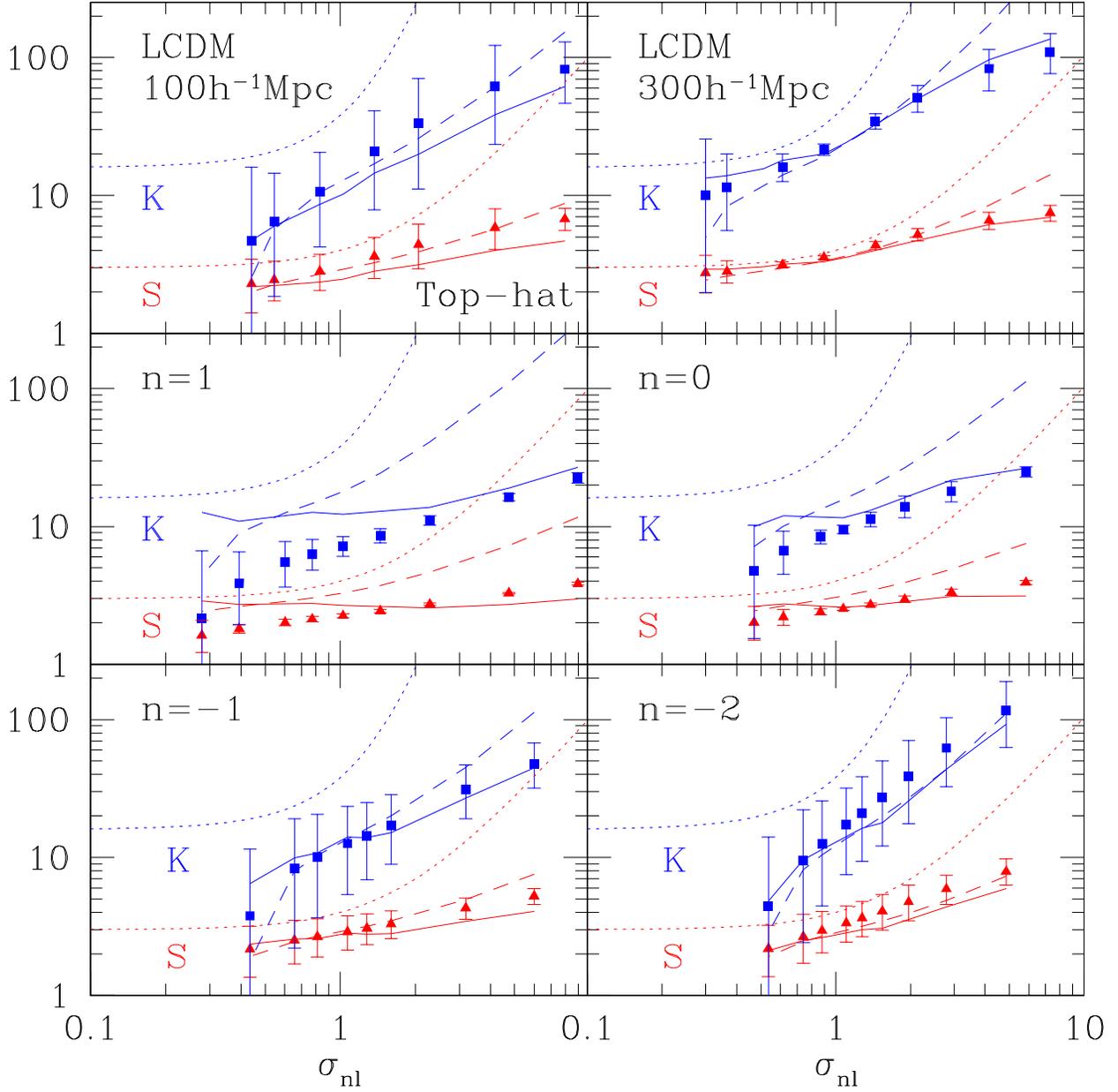} \figcaption{Normalized skewness $S$ and normalized
kurtosis $K$ against $\sigma_{\rm nl}$ from simulations and the
log-normal PDF predictions.  The symbols represent the values estimated
from simulations (the quoted $1\sigma$ error-bars represent the scatter
in the realizations). The meaning of predictions plotted in different
lines is explained in the text.  Top-hat smoothing is
assumed.\label{fig:hiera}}
\end{figure}

To examine the validity of the log-normal PDF more quantitatively, we
compare the normalized skewness $S\equiv
\langle\delta^3\rangle/\langle\delta^2\rangle^2$ and the normalized
kurtosis $K \equiv
(\langle\delta^4\rangle-3\langle\delta^2\rangle^2)/\langle\delta^2\rangle^3$.
The log-normal PDF predicts that
\begin{eqnarray}
 S(R) &=& 3 + \sigma_{\rm nl}^2(R), 
\label{eq:skew}\\
 K(R) &=& 16 + 15\sigma_{\rm nl}^2(R) + 6\sigma_{\rm nl}^4(R) 
   + \sigma_{\rm nl}^6(R) , 
\label{eq:kurt}
\end{eqnarray} 
which are plotted in dotted lines in Figure\ref{fig:hiera} for six
models; two LCDM models with $L_{\rm box}=100h^{-1}$Mpc and
$300h^{-1}$Mpc, and four scale-free models with $n=1$, 0, -1, and $-2$.

In practice, however, the density field $\delta$ in numerical
simulations does not extend the entire range between $-1$ and $\infty$,
but rather is limited as $\delta_{\rm min}<\delta<\delta_{\rm max}$ due to
the finite size of the simulation box. Thus the $n$-th order moments of
$\delta$ in simulations may be better related to
\begin{equation}
\label{eq:moment_with_cutoff}
 \langle\delta^n\rangle' = \int_{\delta_{\rm min}}^{\delta_{\rm
 max}}\delta^nP_{\rm LN}^{(1)}(\delta)d\delta .
\end{equation}
The specific values for $\delta_{\rm min}$ and $\delta_{\rm max}$ may be
roughly estimated from the condition that the expectation number of
independent sampling spheres in the simulation box for
$\delta<\delta_{\rm min}$ or $\delta>\delta_{\rm max}$ becomes unity:
\begin{eqnarray}
 \frac{L_{\rm box}^3}{4\pi R^3/3}\int_{\delta_{\rm max}}^\infty 
P_{\rm LN}^{(1)}(\delta)d\delta &=& 1, \label{eq:cri_max}\\
 \frac{L_{\rm box}^3}{4\pi R^3/3}\int_{-1}^{\delta_{\rm min}}
P_{\rm LN}^{(1)}(\delta)d\delta &=& 1. \label{eq:cri_min}
\end{eqnarray}
Dashed lines in Figure \ref{fig:hiera} show the log-normal PDF
predictions based on equations (\ref{eq:moment_with_cutoff}) to
(\ref{eq:cri_min}). The filled triangles and squares represent the
measurement of $S$ and $K$ from the simulations, and finally the solid
lines indicate the log-normal PDF predictions using equation
(\ref{eq:moment_with_cutoff}) with the actual values for $\delta_{\rm
min}$ and $\delta_{\rm max}$ in the simulations.  Except for the $n=1$
scale-free model, the predictions in solid lines reproduce the
simulation data very well, which indicates that the log-normal PDF is in
fact a good approximation.  The relatively large discrepancy between the
log-normal prediction and the simulation in the $n=1$ model is real
since one can clearly recognize the systematic tendency with respect to
$n$; models with smaller $n$, i.e., with substantial power on large
scales , are better described by the log-normal PDF.  This is consistent
with the discussion by Bernardeau (1994).

Incidentally both the current simulations and the log-normal PDF
approximation confirmed the relatively strong scale-dependence of $S$
and $K$ for $\sigma_{\rm nl} >1$ as pointed out earlier by Lahav et
al. (1993) and Suto (1993). In fact the degree of their scale-dependence
is also sensitive to the underlying power spectrum.  Thus the
hierarchical ansatz for the higher-order clustering is not valid in
general.

In summary, we find that the one-point log-normal PDF remains a fairly
accurate model for the cosmological density distribution even up to
$\sigma_{\rm nl} \sim 4$ and $\delta \sim 100$, fairly independently of
the shape of the underlying power spectrum of density fluctuations.  The
range of validity turns out to be significantly broader than those from
the previous studies based on lower-resolution simulations;
$0.1\simlt\sigma\simlt 0.6$ and $\delta\simlt 4$ (Kofman et al. 1994),
and $0.3\simlt\sigma\simlt 1.5$ and $\delta\simlt9$ (Bernardeau \&
Kofman 1995), for instance.

\subsection{The two-point PDF}
\label{subsec:2pcompare}

While we would like to perform the similar comparison for the two-point
PDFs, it is a function of four variables, $\delta_1$, $\delta_2$, $R$,
and $r$, and thus the comparison becomes rather complicated. Therefore
we decided to use the conditional two-point PDF for the purpose:
\begin{eqnarray}
 P^{(2)}(\delta_1|\delta_2; r) \equiv
 \frac{P^{(2)}(\delta_1, \delta_2; r)}{P^{(1)}(\delta_2)} .
\end{eqnarray}
Since we already made sure that the one-point PDF is very accurately
approximated by the log-normal, our task is to see if simulation
results fit the conditional two-point log-normal PDF:
\begin{eqnarray}
 P^{(2)}_{\rm LN}(\delta_1|\delta_2; r)
 \equiv
 \sqrt{\frac{S}{2\pi(S^2-X^2)}}
 \exp\left[-\frac{(S L_1-X L_2)^2}{2S(S^2-X^2)}
 \right]\frac{1}{(1+\delta_1)} . \label{eq:con2pLN}
\end{eqnarray}

The evaluation of the conditional two-point PDFs from simulations is
carried out as follows.  From the smoothed density fields computed on
the $512^3$ grid points, we first select those grid points with
$\delta_2-\Delta\delta_2/2 < \delta < \delta_2+\Delta\delta_2/2$.  The
bin size $\Delta\delta_2$ is adjusted for each value of $\delta_2$ so
that approximately $10^5$ grid points satisfy the condition.  Next we
pick up all grid points separated at $r-\Delta r/2\sim r+\Delta r/2$
from the above grids.  The separation interval $\Delta r$ is chosen so
that $4\pi r^2 \Delta r (512/L_{\rm box})^3 \sim 10^3$.  Finally we
compute the conditional two-point PDF with a constant bin size of 0.1
in $\delta_1$.

\begin{figure}
\plotone{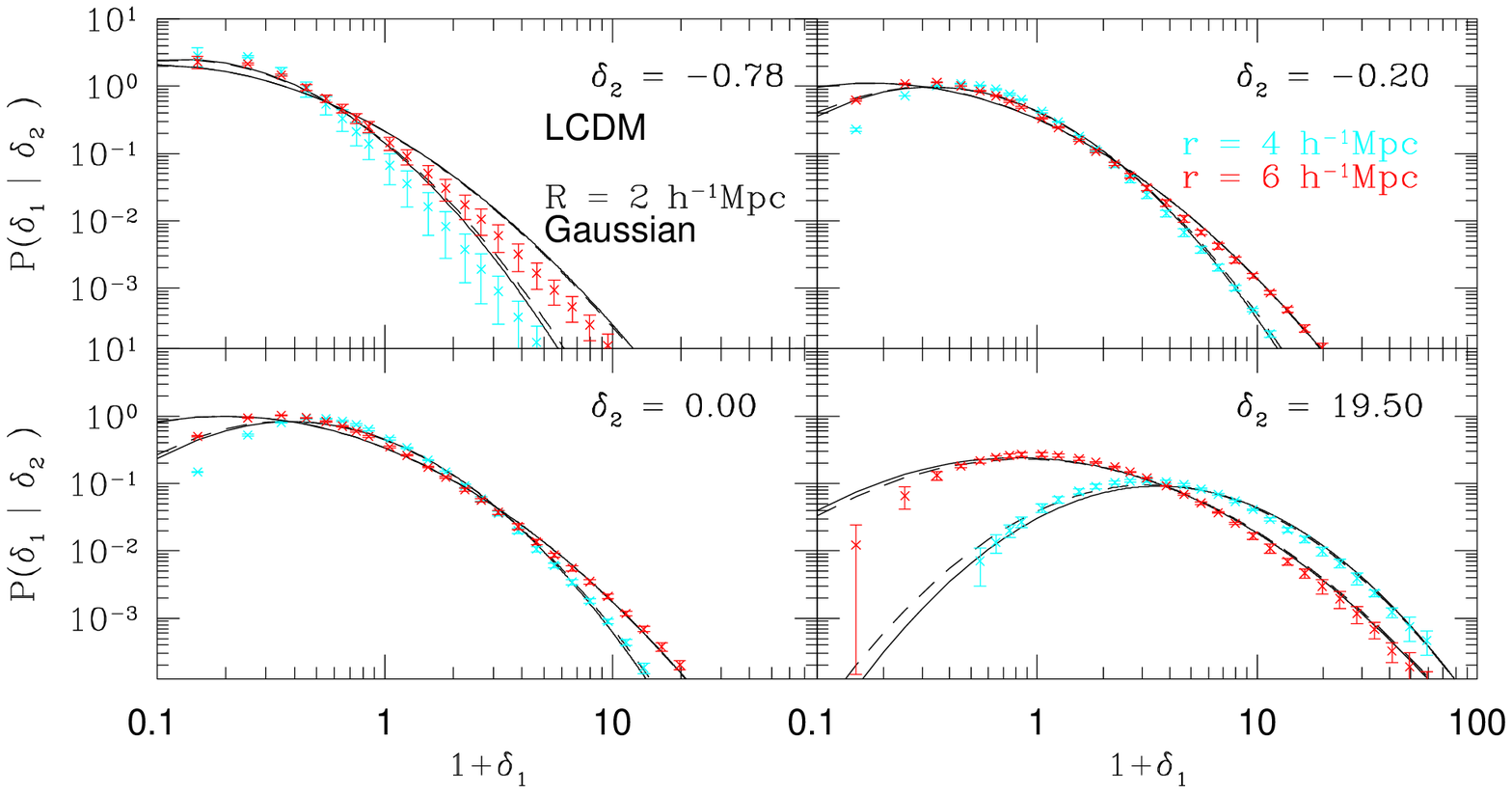} 
\plotone{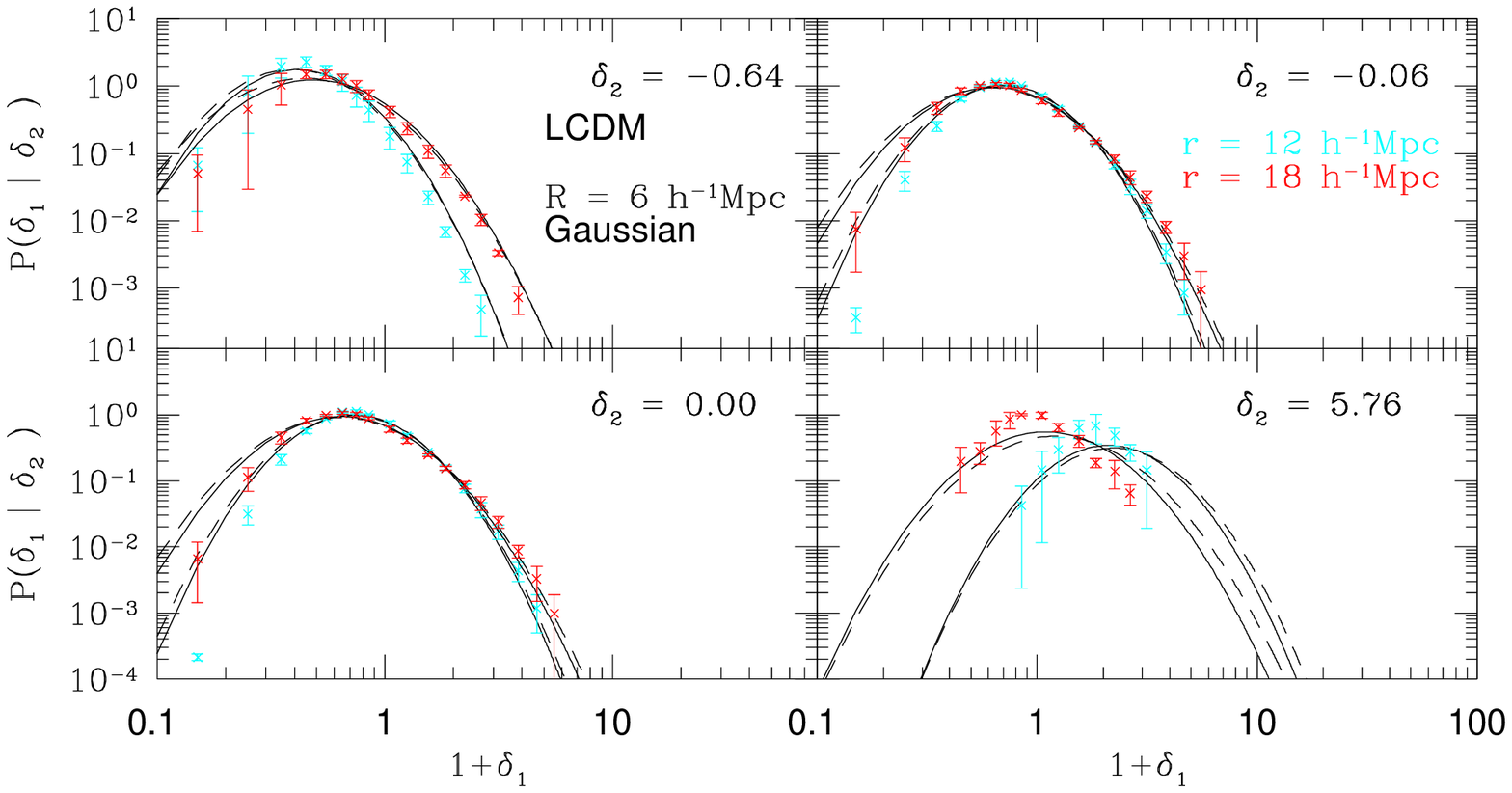} 
\caption{Two-point PDFs in LCDM model with Gaussian smoothing over
$R=2$\himpc ({\it upper} four panels) and $R=6$\himpc ({\it lower} four
panels).  The results at separation $r=2R$ and $3R$ are plotted.
The {\it solid} and {\it long-dashed} lines represent the
log-normal PDF adopting $\sigma_{\rm nl}$ and $\xi_{\rm nl}$ calculated
directly from the simulations and estimated from the nonlinear fitting
formula of Peacock \& Dodds (1996), respectively.  \label{fig:2pLN}}
\end{figure}

Figure \ref{fig:2pLN} plots the resulting PDFs for the LCDM model with
the Gaussian smoothing window; the upper four panels show the PDFs for
the separation $r=4$ and $6h^{-1}{\rm Mpc}$ and the smoothing length
$R=2 h^{-1}{\rm Mpc}$, while the lower four panels for $r=12$ and
$18h^{-1}{\rm Mpc}$ and $R=6 h^{-1}{\rm Mpc}$. Solid lines indicate
the conditional log-normal PDFs adopting the values of $\sigma_{\rm
nl}$ and $\xi_{\rm nl}$ from simulations, while dashed lines show
those using the PD predictions (Tables \ref{tbl:sigma_cdm} to
\ref{tbl:corr_sf}). Clearly the log-normal PDF is a reasonably good
approximation. The deviation at $\delta_2 \simlt -0.5$, on the other
hand, seems real and may be an enhanced feature that we noted in the
one-point PDF (Fig. \ref{fig:1pLN_CDM}).

\begin{figure}
\plotone{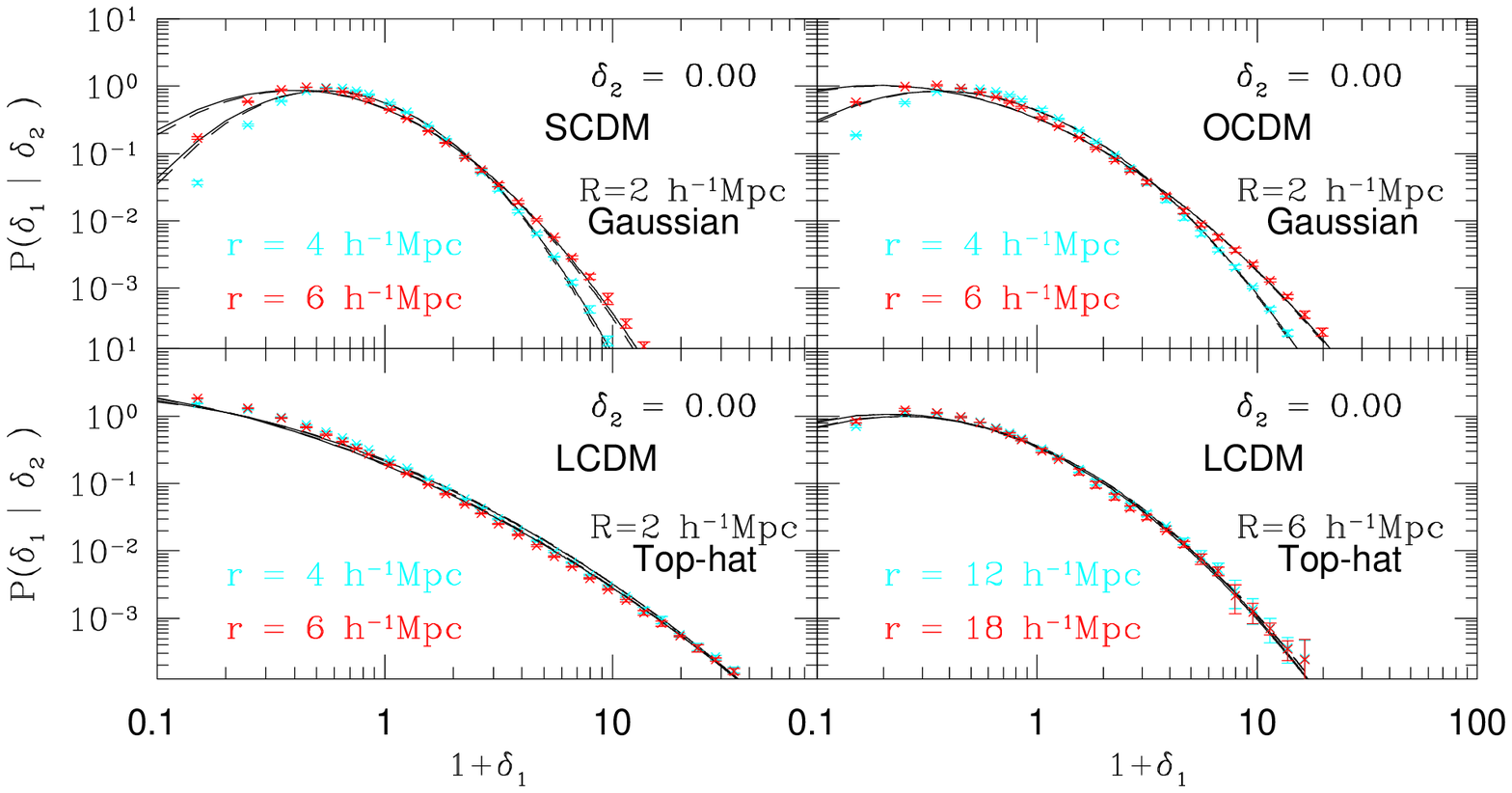}
\plotone{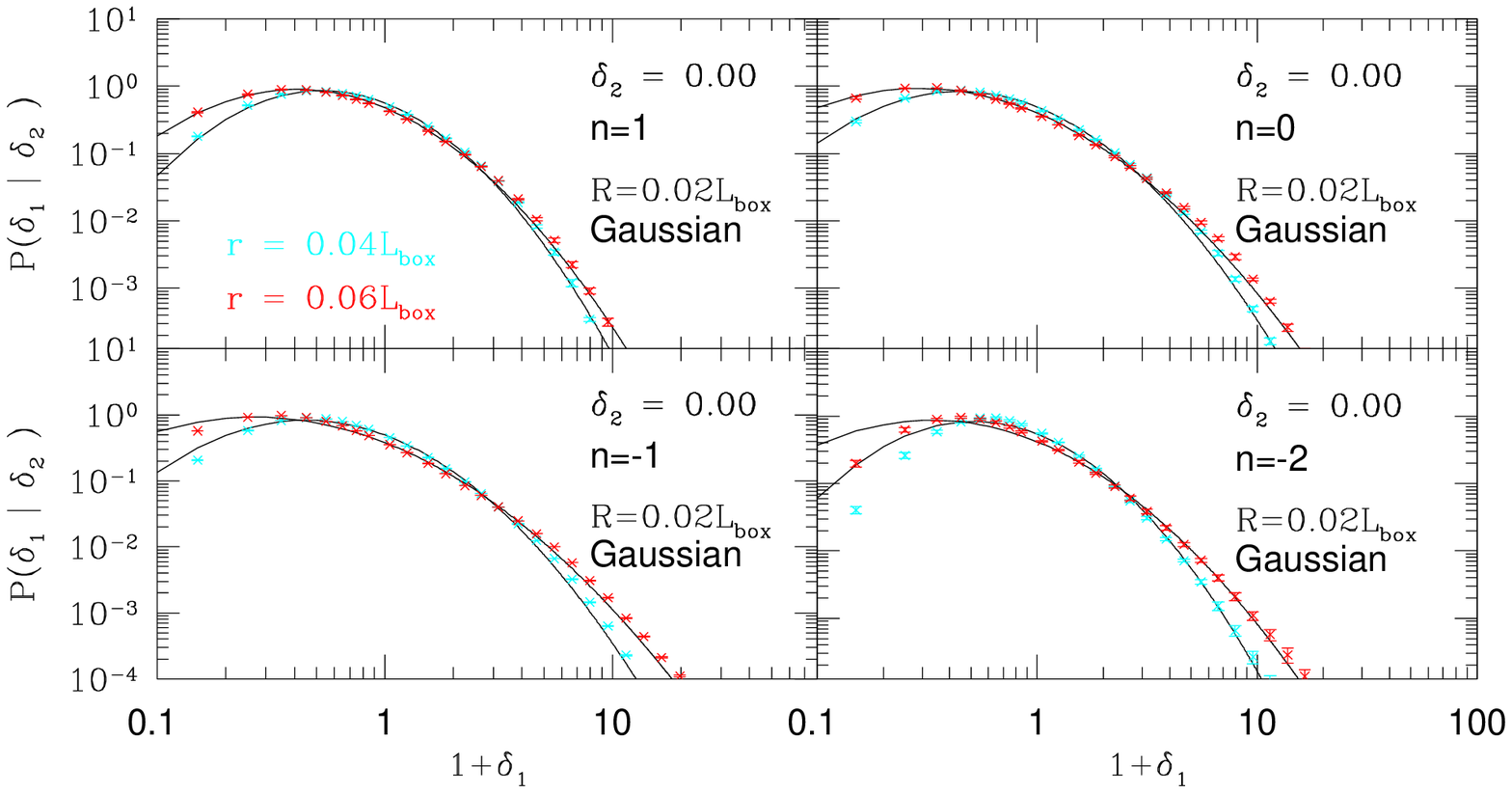}
\figcaption{Two-point PDFs for different models and smoothing window
 functions. The {\it upper} four panels plot the results in CDM models,
 while the {\it lower} four panels in scale-free models.
 \label{fig:2pLN_others}}
\end{figure}

\begin{figure}
\plotone{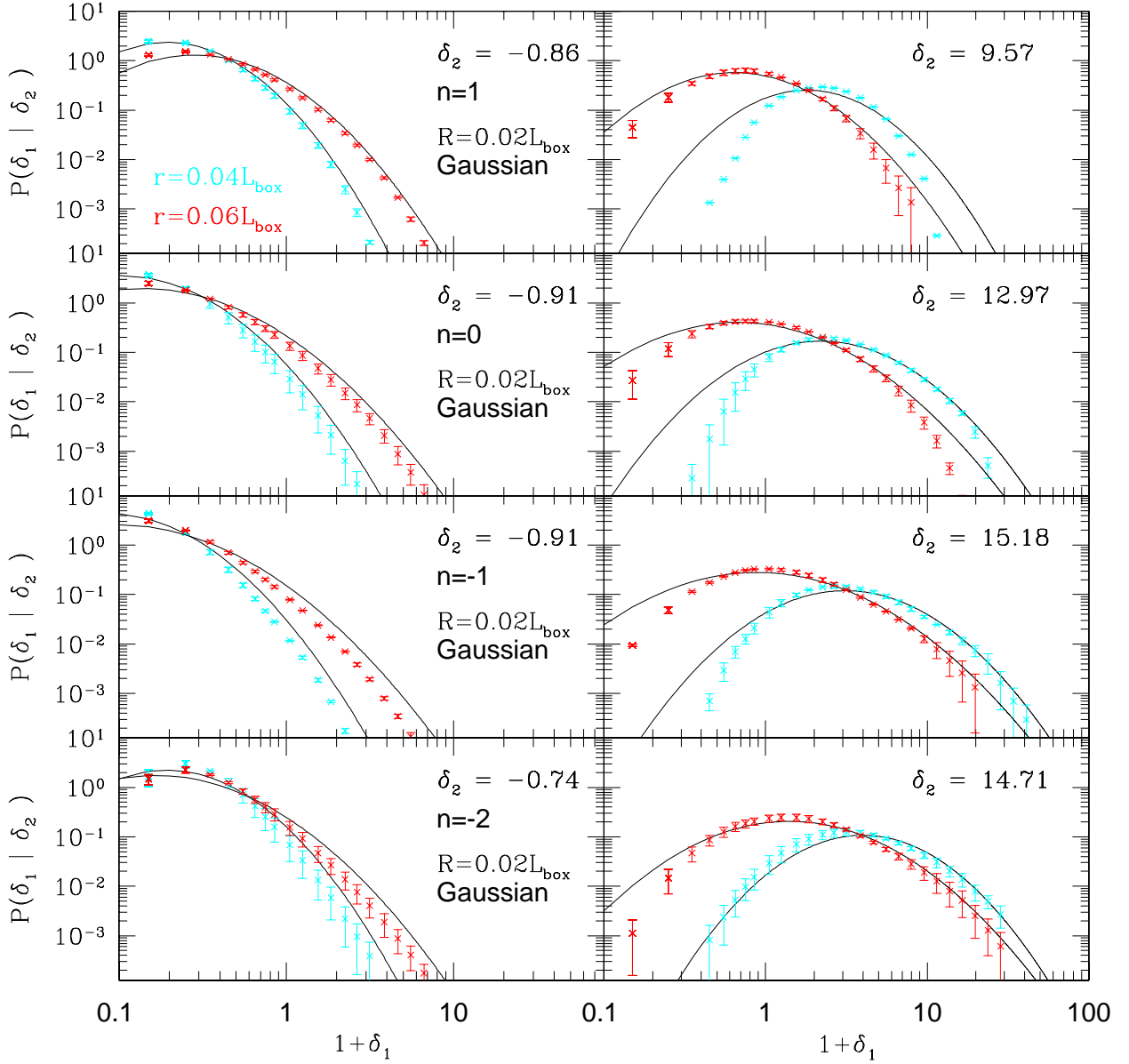}
\figcaption{Two-point PDFs for the scale-free models at negative ({\it
left}) and positive ({\it right}) tails of the distribution of
$\delta_2$.
 \label{fig:2pLN_lh}}
\end{figure}

Figure \ref{fig:2pLN_others} indicates that the good agreement is
achieved not only in the Gaussian smoothed LCDM model, but also can be
found in the other models and/or the top-hat smoothing.  Figure
\ref{fig:2pLN_lh} plots the conditional two-point PDFs at $\delta_2
\simlt - 0.7$ and $\delta_2 \simgt 10$ in the scale-free models, where
the deviation from the log-normal becomes manifest.
Considering the error-bars estimated from the different realizations for
each model, the deviation seems statistically real.

As in the case of the one-point PDF, we illustrate the validity of the
two-point log-normal PDF using the moments.  Specifically we evaluate
$\langle (\delta_1\delta_2)^2\rangle$ and
$\langle(\delta_1\delta_2)^3\rangle$ according to
\begin{eqnarray}
 \langle(\delta_1\delta_2)^n\rangle (r)
\equiv\iint_{{\cal C}(\delta_1, \delta_2)}
 (\delta_1\delta_2)^n 
 P_{\rm LN}^{(2)}(\delta_1, \delta_2; r)d\delta_1 d\delta_2 ,
\end{eqnarray}
where we select the range of the integration as
\begin{equation}
 {\cal C}(\delta_1, \delta_2) 
= \{(\delta_1, \delta_2) | 
\delta_{\rm min}\le \delta_1 \le \delta_{\rm max}, 
\delta_{\rm min}\le \delta_2 \le \delta_{\rm max}\},
\end{equation}
from the values of $\delta_{\rm min}$ and $\delta_{\rm max}$ directly
measured from each simulation model.  The results are summarized in
Table \ref{tbl:hiera2}, which indicates again that the two-point
log-normal PDF predictions reproduce the simulation data except for
$n\ge 0$.  Thus we also conclude that the two-point log-normal model
describes fairly well the PDF of cosmological fluctuations for most
regions of $\delta_1$ and $\delta_2$ of interest; the small but finite
deviations exist only in $\delta_1 \simlt -0.5$ and/or $\delta_2 \simlt
-0.5$, and in $\delta_1 \simgt 10$ and $\delta_2\simgt 10$.

\subsection{Does the log-normal transformation approximate 
the gravitational evolution of the density fluctuations ?}
\label{subsec:crue}

The agreement between the log-normal predictions and the simulation
results might be interpreted as indirect evidence that the log-normal
transformation (eq.[\ref{eq:trans1}]) is a good approximation for
the nonlinear gravitational growth of the cosmological density
fluctuations, at least on average.

In order to see if this is really the case, we consider the relation of
the smoothed density fields at the same comoving position but at
different redshifts $z_1$ and $z_2$.  For this purpose, we use one
realization from the LCDM model evolved in $L_{\rm box}=300\himpc$.  If
the log-normal transformation (\ref{eq:trans1}) is exact, the density
fluctuations, $\delta(z_1)$ and $\delta(z_2)$, should satisfy
\begin{equation}
 1+\delta(z_2)   =
  \frac{1}{\sqrt{1+\sigma^2_{\rm nl}(z_2)}}
  \exp\left[\frac{\sqrt{\ln\{1+\sigma^2_{\rm nl}(z_2)\}}}
       {\sqrt{\ln\{1+\sigma^2_{\rm nl}(z_1)\}}}
       \ln\left\{\sqrt{
	   1+\sigma^2_{\rm nl}(z_1)}(1+\delta(z_1))\right\}\right].
       \label{eq:transLN}
\end{equation}

\begin{figure}
\plotone{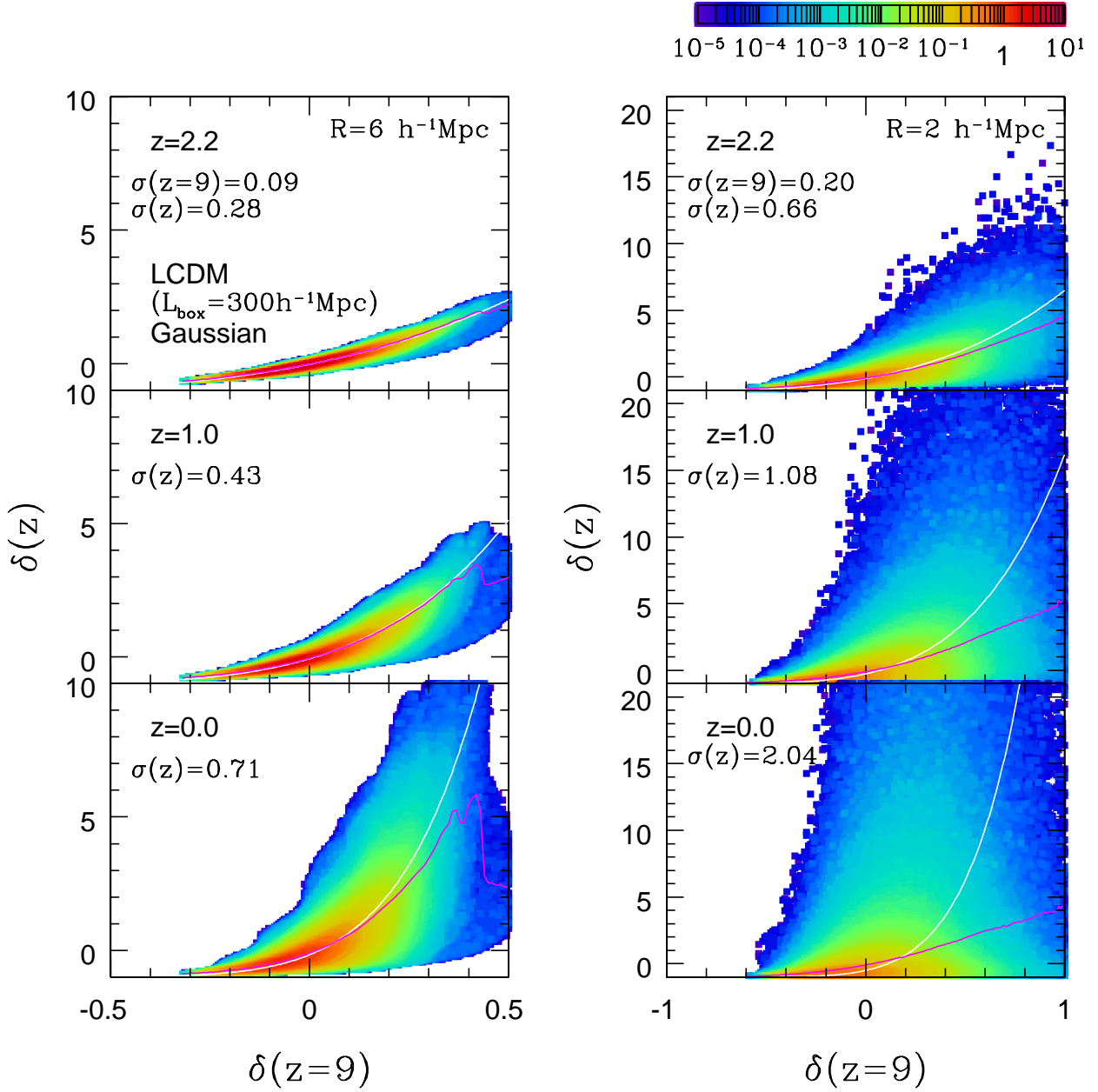} 
\figcaption{Contour plots of the joint probability
$P(\delta(z)~|~\delta(z=9))$ in the LCDM model ($L_{\rm box}=300\himpc$)
 with Gaussian smoothing window; $R=6 h^{-1}$Mpc ({\it right}) and
 $R=2 h^{-1}$Mpc ({\it right}).
 The top, middle and bottom panels correspond to correlations of $\delta
 (z=2.2)$, $\delta (z=1)$, and $\delta (z=0)$ against $\delta(z=9)$,
 respectively.  The {\it white} lines represent the log-normal
 transformation (\ref{eq:transLN}) and the {\it magenta} lines are the
 conditional mean at a fixed $\delta(z=9)$.  \label{fig:devo_con}}
\end{figure}

Figure \ref{fig:devo_con} plots the color contour of the joint
probability $P(\delta(z), \delta(z=9))$ of densities at $z=2.2$, $1.0$,
and $0$ against that at $z=9$ on the same grid points in the LCDM
model. We adopt the Gaussian smoothing with $R=6 h^{-1}{\rm Mpc}$ ({\it
Left}) and $2 h^{-1}{\rm Mpc}$ ({\it Right}).  The solid lines in white
and magenta represent the the log-normal transformation
(\ref{eq:transLN}) and the conditional mean from simulations for a fixed
$\delta(z=9)$.  The log-normal transformation traces the mean relation
of simulations to some extent only when the nonlinearity is weak (see,
higher $z$ and larger $R$ cases).  On the other hand, in the nonlinear
region, the transformation (\ref{eq:transLN}) starts to deviate from the
mean relation of the simulations significantly, and the distribution
around the mean relation becomes broad.  The similar tendency was found
in a somewhat different analysis by Coles, Melott, \& Shandarin (1993).
In a sense this is a physically natural and expected result, but then it
makes even more difficult to account for the good agreement between the
log-normal and simulation PDFs in those scales.

\begin{figure}
\plotone{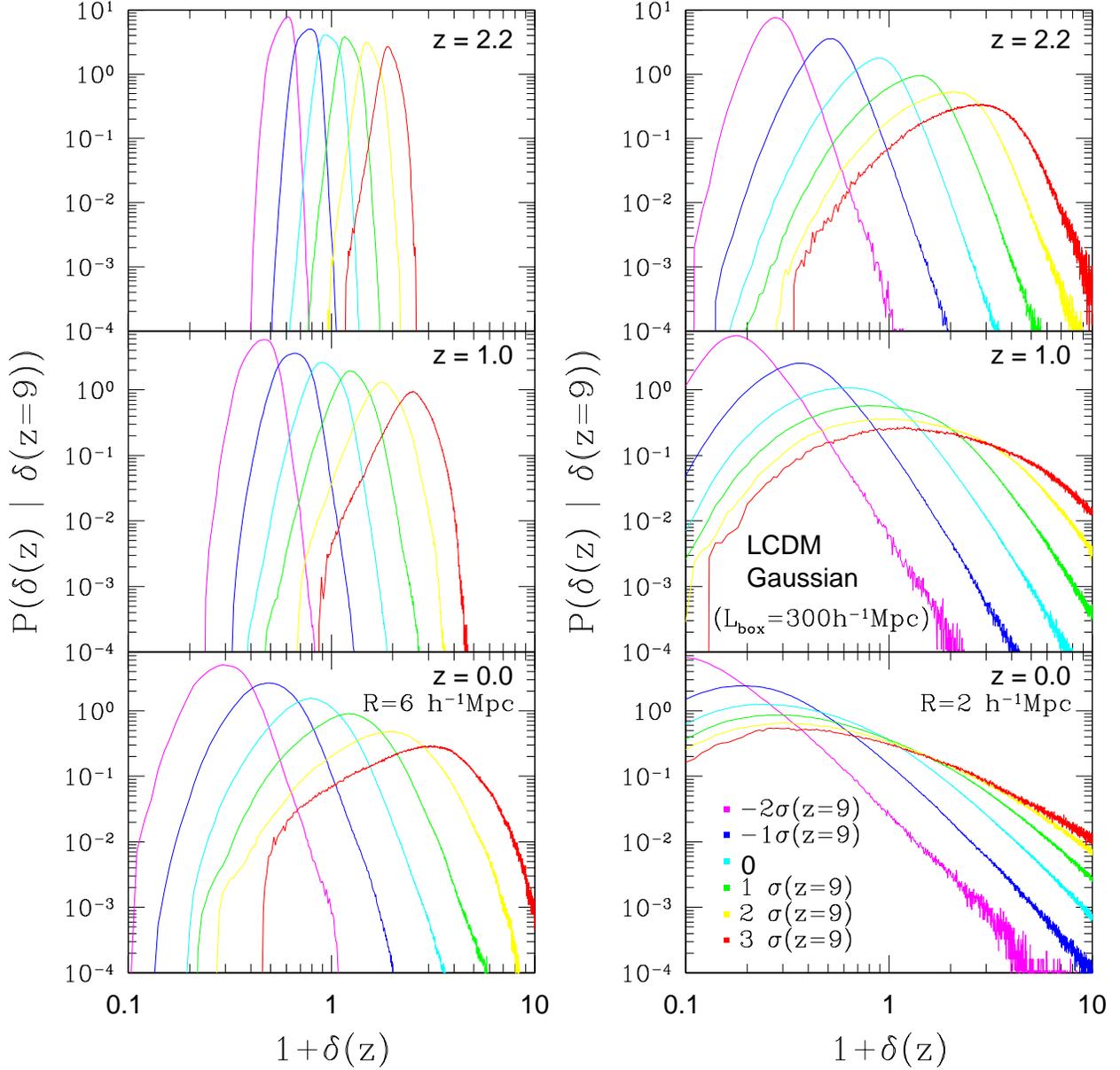}
\figcaption{Conditional probability $P(\delta(z)~|~\delta(z=9))$ for a
 fixed $\delta(z=9)$ corresponding to each panel of
 Fig. \ref{fig:devo_con}. In each panel, results for $\delta(z=9) = -2$,
 $-1$, 0, 1, 2, and 3 times the $\sigma(z=9)$, the rms of $\delta(z=9)$,
 are plotted.  \label{fig:devo_slice}}
\end{figure}

To better understand the distribution of the linear and its evolved
density fields, we compute the conditional probability $P(\delta(z)
~|~ \delta(z=9))$, i.e., the slice of Figure \ref{fig:devo_con} at a
given $\delta(z=9)$.  The results are plotted in Figure
\ref{fig:devo_slice} which exhibits a some regularity in the
distribution. The peak positions seem to show some scaling with
respect to the value of $\delta(z=9)$, and also the tail of the
distribution asymptotically approaches a single power-law.  While we
do not yet fully understand the behavior, this regularity in the
distribution function may be useful in explaining our findings that
the one-point and two-point log-normal PDFs work well empirically.

\section{CONCLUSIONS AND DISCUSSION}
\label{sec:conclusion}

In the present paper, we have estimated the probability distribution
functions of cosmological density fluctuations from the
high-resolution N-body simulations with the Gaussian initial
condition.  In particular, we have critically examined the validity of
the log-normal models for the one- and two-point PDFs both in weakly
and in strongly nonlinear regimes.

We have shown that the one-point log-normal PDF is a fairly accurate
model not only in a weakly nonlinear regimes as claimed previously, but
also in more strongly nonlinear regimes even up to $\sigma_{\rm nl} \sim
4$ and $\delta \sim 100$.  Furthermore, we extended the analysis to the
two-point PDF, and found that the log-normal PDF serves also as an
empirically accurate model for the range of densities of interest.  This
is the case fairly independently of the shape of the underlying power
spectrum of density fluctuations, although models with large power on
small scales (e.g., $n \ge 0$ scale-free models in our examples) seem to
show a small deviation from the log-normal prediction at the tails of
the distribution, especially for $\delta \simlt -0.5$.  In particular,
the log-normal PDF reproduces very well the skewness and kurtosis
measured from the simulation data, when the finite size of the
simulation volume is properly taken into account.

The degree of agreement of the log-normal models that we have shown is
amazing considering the fact that the underlying mapping between the
initial and the evolved density fields differs significantly from the
simulation results even in an averaged sense. We have explicitly shown
the probability distribution of the initial and the evolved density
fields from simulations, although we were not able to provide a
physical explanation for the origin of the log-normal PDF. This should
be left as our future work and we would like to come back later
elsewhere. For this purpose, other theoretical approaches based on
perturbation theory (Bernardeau 1992, 1994) and the spherical collapse
model (Fosalba \& Gazta\~{n}aga 1998) may be helpful.

Nevertheless our present work provides an empirical justification for
the use of the log-normal PDF in a variety of theoretical model
predictions. For instance, Matsubara \& Yokoyama (1996) proposed to
evaluate the effect of the nonlinear gravitational evolution on the
genus statistics using the log-normal mapping.  Taruya \& Suto (2000)
constructed an analytical model for halo biasing on the basis of the
one-point log-normal PDF of underlying mass density field.  Hikage,
Taruya, \& Suto (2001) applied this biasing model in their predictions
of the genus for clusters of galaxies.

Finally, the present results might be useful in considering the
prediction of weak lensing statistics (Valageas 2000; Munshi \& Jain
2000).
To construct a model for PDF in redshift space is another
important topic (e.g., Watts \& Taylor 2001; Hui, Kofman \& Shandarin
2000), which is relevant in discussing Ly-$\alpha$ forests
(Gazta\~{n}aga \& Croft 1999).

\bigskip
\bigskip

We thank Y. P. Jing for kindly providing us his N-body data, and an
anonymous referee for constructive comments. I.K. and A.T. gratefully
acknowledge support from Takenaka-Ikueikai fellowship, and a JSPS (Japan
Society for the Promotion of Science) fellowship, respectively.  
Numerical computations were
carried out at RESCEU (Research Center for the Early Universe,
University of Tokyo), ADAC (the Astronomical Data Analysis Center) of
the National Astronomical Observatory, Japan, and at KEK (High Energy
Accelerator Research Organization, Japan).  This research was supported
in part by the Grant-in-Aid by the Ministry of Education, Science,
Sports and Culture of Japan (07CE2002, 12640231) and by the
Supercomputer Project (No.00-63) of KEK.



\clearpage

\begin{deluxetable}{lcccccc}
\tablecolumns{7}
\tablewidth{0pc}
\tablecaption{Simulation parameters for the CDM models.
\label{tbl:simpara}}
\tablehead{
\colhead{Model} & \colhead{$\Omega_0$} &  \colhead{$\lambda_0$}
& \colhead{$\Gamma$\tablenotemark{\dagger}} & \colhead{$\sigma_8$}
& \colhead{$L_{\rm box}$[$h^{-1}$Mpc]} & \colhead{realizations}}
\startdata
SCDM & 1.0 & 0.0 & 0.50 & 0.6 & 100 & 3  \\
LCDM & 0.3 & 0.7 & 0.21 & 1.0 & 100 & 3  \\
LCDM300 & 0.3 & 0.7 & 0.21 & 1.0 & 300 & 3  \\
OCDM & 0.3 & 0.0 & 0.25 & 1.0 & 100 & 3  \\
\enddata
\tablenotetext{\dagger}{Shape parameter of the power spectrum.}
\end{deluxetable}

\begin{deluxetable}{lcccc}
\tablecolumns{5}
\tablewidth{0pc}
\tablecaption{Amplitude of $\sigma_{\rm nl}(R)$ evaluated from
the CDM simulations. The values in parenthesis denote those estimated
 from the nonlinear fitting formula of Peacock \& Dodds (1996). 
\label{tbl:sigma_cdm}}
\tablehead{
  smoothing & $R$ [$h^{-1}$Mpc] & SCDM & LCDM & OCDM}
\startdata
         & 2 & 2.33 (2.24) & 4.17 (4.08) & 4.37 (4.23) \\
 top-hat & 6 & 0.79 (0.77) & 1.37 (1.40) & 1.37 (1.38) \\
         & 18 & 0.23 (0.24) & 0.44 (0.50) & 0.43 (0.47) \\
\hline
          & 2 & 1.11 (1.08) & 1.95 (1.96) & 1.97 (1.96) \\
 Gaussian & 6 & 0.36 (0.35) & 0.64 (0.69) & 0.63 (0.67) \\
          & 18& 0.065 (0.090) & 0.15 (0.22) & 0.13 (0.21) \\
 \enddata
\end{deluxetable}

\begin{deluxetable}{lccccc}
\tablecolumns{6}
\tablewidth{0pc}
\tablecaption{Rms $\sigma_{\rm nl}(R)$ in the scale-free
 simulations. \label{tbl:sigma_sf}}
\tablehead{
  smoothing & $R$ [$L_{\rm box}$] & $n=1$ & $n=0$ & $n=-1$ & $n=-2$}
\startdata
          & 0.02 & 2.48 & 3.10 & 3.18 & 2.79\\
 top-hat  & 0.05 & 0.80 & 1.10 & 1.28 & 1.28\\
          & 0.15 & 0.15 & 0.28 & 0.43 & 0.54\\
\hline
          & 0.02 & 1.00 & 1.34 & 1.51 & 1.47\\
 Gaussian & 0.05 & 0.26 & 0.44 & 0.62 & 0.71\\
          & 0.15 & 0.03 & 0.09 & 0.17 & 0.25\\
 \enddata
\end{deluxetable}

\begin{deluxetable}{ccccc}
\tablecolumns{5}
\tablewidth{0pc}
\tablecaption{Amplitude of $\xi_{\rm nl}(r; R)$ evaluated from the CDM
 simulations with Gaussian smoothing. 
 The values in parenthesis denote those estimated from the
 nonlinear fitting formula of Peacock \& Dodds (1996).\label{tbl:corr_cdm}}
\tablehead{
  $R$ [$h^{-1}$Mpc] & $r$ [$h^{-1}$Mpc] & SCDM & LCDM & OCDM}
\startdata
 2 & 4 & 0.68 (0.64) & 2.15 (2.10) & 2.05 (2.06) \\
 2 & 6 & 0.36 (0.35) & 1.10 (1.15) & 1.06 (1.10) \\
\hline
 6 & 12& 0.058 (0.063) & 0.21 (0.26) & 0.20 (0.24) \\
 6 & 18& 0.021 (0.028) & 0.10 (0.144) & 0.087 (0.127) \\
\enddata
\end{deluxetable}

\begin{deluxetable}{cccccc}
\tablecolumns{6}
\tablewidth{0pc}
\tablecaption{Two-point correlation $\xi_{\rm nl}(r; R)$ in the
 scale-free simulations with Gaussian smoothing.\label{tbl:corr_sf}}
\tablehead{
  $R$ [$L_{\rm box}$] & $r$ [$L_{\rm box}$] & $n=1$ & $n=0$ & $n=-1$ & $n=-2$}
\startdata
 0.02 & 0.04 & 0.40 & 0.82 & 1.29 & 1.35\\
 0.02 & 0.06 & 0.13 & 0.36 & 0.67 & 0.83\\
\enddata
\end{deluxetable}

\begin{deluxetable}{ccccc}
\tablecolumns{5}
\tablewidth{0pc}
\tablecaption{Two-point moments, $\langle(\delta_1\delta_2)^2\rangle$ and
 $\langle(\delta_1\delta_2)^3\rangle$, from simulations and the
 log-normal PDF predictions.
The LCDM models adopt the Gaussian smoothing with $R=2\himpc$ and
 the moments are evaluated at the pair-separation of  $r=4\himpc$.
For the scale-free models, $R=0.02L_{\rm box}$ and $r=0.04L_{\rm box}$.
 \label{tbl:hiera2}}
\tablehead{
 Model & $\langle(\delta_1\delta_2)^2\rangle$ &
 $\langle(\delta_1\delta_2)^3\rangle$ & $\delta_{\rm min}$ & $\delta_{\rm max}$}
\startdata
 LCDM & $(0.76\pm 0.4)\times 10^3$ & $(1.5\pm 1.4)\times 10^6$
 & \nodata & \nodata\\
 log-normal      & $0.64\times 10^3$ & $1.1\times 10^6$ & $-0.96$ & 90\\
\hline
 LCDM300 & $(1.4\pm 0.2)\times 10^3$ & $(4.9\pm 1.5)\times 10^6$
 & \nodata & \nodata\\
 log-normal      & $1.3\times 10^3$ & $6.2\times 10^6$ & $-0.98$ & 165\\
\hline
 $n=1$ & $3.1\pm 0.1$ & $73\pm 11$ & \nodata & \nodata\\
 log-normal & 4.0 & 140 & $-0.98$ & 14\\
\hline
 $n=0$ & $24\pm 3$ & $(2.6\pm 0.6)\times 10^3$ & \nodata & \nodata\\
 log-normal & 30 & $5.6\times 10^3$ & $-0.99$ & 30\\
\hline
 $n=-1$ & $88\pm 18$ & $(2.2\pm 1.0)\times 10^4$ & \nodata & \nodata\\
 log-normal & 102 & $3.9\times 10^4$ & $-0.98$ & 42 \\
\hline
 $n=-2$ & $120\pm 40$ & $(4.1\pm 2.1)\times 10^4$ & \nodata & \nodata\\
 log-normal & 120 & $5.3\times 10^4$ & $-0.96$ & 46 
\enddata
\end{deluxetable}

\clearpage

\end{document}